\documentstyle[preprint,aps]{revtex}
\newcommand{\albtwo}{$\rm AlB_2$}
\newcommand{\mgbtwo}{MgB$_2$}
\newcommand{\mgalbtwo}{$\rm Mg_{0.5}Al_{0.5}B_2$}
\newcommand{\casitwo}{CaSi$_2$}
\newcommand{\cabesitwo}{$\rm CaBe_{x}Si_{2-x}$}
\newcommand{\cabesi}{$\rm CaBeSi$}
\newcommand{\bebtwo}{BeB$_2$}
\begin{document}
\draft

\title{Electronic and structural properties of  superconducting
MgB$_2$,  CaSi$_2$ and  related compounds}

\author{G. Satta$^{\dag}$, G. Profeta$^*$,  
 F. Bernardini$^{\dag}$, A. Continenza$^*$ and S. Massidda$^{\dag}$ }
\address{$^*$Istituto Nazionale di Fisica della Materia (INFM) and
Dipartimento di Fisica,
Universit\`a degli Studi di L'Aquila, I--67010 Coppito (L'Aquila),
Italy }
\address{$^{\dag}$Istituto Nazionale di Fisica della Materia (INFM)  and
Dipartimento di Fisica  Universit\`a degli
Studi di Cagliari,  09124 Cagliari, Italy }

\maketitle

\begin{abstract}
We report a detailed study of the electronic and structural 
properties of the 39K superconductor \mgbtwo\ and of several related 
systems of the same family, namely \mgalbtwo, \bebtwo, \casitwo\ and 
\cabesi.  Our calculations, which include zone-center phonon frequencies
and transport properties, are performed within the local density
approximation  to the density functional theory, using the
full-potential linearized augmented plane wave (FLAPW) and the
norm-conserving pseudopotential methods.   Our results indicate
essentially three-dimensional properties for these compounds; however,
strongly two-dimensional $\sigma$-bonding bands 
contribute significantly at the Fermi level. Similarities and 
differences between \mgbtwo\ and \bebtwo\ (whose superconducting properties
have not been yet investigated) 
are analyzed in detail.  Our calculations for
\mgalbtwo\  show that metal substitution cannot be fully described 
in a rigid band model.  \casitwo\ is studied as a function of pressure, and
 Be substitution in the Si planes leads to a stable compound similar
in many aspects to diborides.
\end{abstract}
\newpage
\section{Introduction}

The recent discovery\cite{mgb2exp} of superconductivity at $\approx 39$K
in the simple intermetallic compound  MgB$_2$ is particularly interesting
for many reasons. In the first instance, this critical
temperature is by far the highest if we exclude oxides and C$_{60}$ based
materials.  Furthermore, B isotope effect\cite{mgb2exp2}  suggests
that \mgbtwo\ is a BCS phonon-mediated superconductor,  with $T_c$
above
the commonly accepted limits for phonon-assisted superconductivity.
Another reason of interest is that the  AlB$_2$-type structure
of \mgbtwo\  (which can be viewed as an intercalated graphite structure
with  full occupation of interstitial sites centered in  hexagonal prisms
made of B atoms)
is shared by a large class of compounds (more than one hundred), where
the B site can be occupied by Si or  Ge, but also by Ni, Ga, Cu, Ag, Au, Zn
and Cd, and  Mg can be substituted by  divalent or trivalent $sp$
metals, transition elements or rare earths. Such a variety is therefore
particularly suggestive  of a systematic study of diborides and
structurarly related
compounds. To confirm this interest,
a recent study\cite{cava} showed that superconductivity is supressed by
addition of more than about 30\% of Al atoms.
Pressure effects\cite{mgb2pres} also lead to a depression of superconductivity.

Early studies\cite{diborurith,tupitsin,armstrong} of the electronic
structure of diborides by the OPW  and
tight-binding methods have shown similarity with graphite bands. A  more
systematic study of \albtwo--type compounds\cite{mass,mass1}  by the
full-potential
LAPW method showed  in simple metal diborides the presence of interlayer
states, similar to those previously found in
graphite\cite{graphite,posternak}, in
addition to graphite--like $\sigma$ and $\pi$ bands. Transition metal
diborides, on the other hand, show a more complex band structure due
to the metal $d$ states.
A very recent electronic structure calculation\cite{kortus}  on
\mgbtwo\   shows  that metallicity is due to B states.
Phonon frequencies and electron-phonon couplings (at the $\Gamma$ point)
lead these
authors to suggest  phonon-based superconductivity. 
Further, new calculations\cite{lib2} 
on related compounds ($\rm LiB_2$ and graphitic-like B) 
addressed the peculiar contributions to conductivity coming from
the interplay of 2D and 3D Fermi surface features.
An alternative  mechanism has been proposed by Hirsch\cite{hirsch},
based on dressed  hole carriers in the B planes, in the 
presence of almost filled bands.

Recently, superconductivity at  $T_c\approx 14$K was observed in 
$\rm CaSi_2$\cite{Affronte}.
In this compound, high  pressure has been  effectively
used to tune both structural and physical properties\cite{Affronte,Sanfilippo}.
In particular, Bordet and co-workers
\cite{Affronte} identified  a new structural phase  transition for {\rm
$\rm CaSi_2$}  from trigonal to hexagonal  $\rm AlB_2$-type structure, with
superconductivity at  $T_{c}=14~K$, one
of the highest transition temperatures known for a silicon based compound.
The occurrence of this structural phase transition indicates that Si shifts
from an $sp^3$ to an $sp^2$-like hybridization, and experiment points to a clear
enhancement of $T_c$  with high pressure\cite{Affronte}.
Many electronic structure studies have been carried out for  the different
phases\cite{Weaver,Fahy,Bisi} of disilicides. $\rm CaSi_2$  in the $\rm AlB_2$
structure  was  studied by Kusakabe et
al.\cite{Kusakabe}, but a final answer on the stability of this phase
under pressure is still lacking.

As previously pointed out, the $\rm AlB_2$  family of compounds provides a wide
playground for the search of new superconductors, and we intend to provide
here a contribution, based on a detailed  study of electronic and structural
properties together with
phonon spectra, for several diborides and for {$\rm CaSi_2$},
in the trigonal and hexagonal structures.
The questions that we want to address are the following:
(i) how do the electronic properties of MX$_2$ depend on the atomic species
M; namely, can rigid band schemes provide the correct trend, or is the metal
effect
mostly mediated by lattice parameter changes (chemical pressure).
(ii) Ref.\onlinecite{kortus}
shows the presence of both B $\pi$ Fermi surfaces (FS) giving rise
to 3-dimensional states, and of  small 2-dimensional B
$\sigma$-bonding pockets.  It is of course interesting to know which of the
FS
is mostly involved in superconductivity.  The trends observed in the band
structures, together with experimental studies, will contribute to
elucidate this point. In the same way, changes in the Fermi level
density of states with chemical composition will possibly correlate with
$T_c$ values. (iii) Pure \casitwo\  shows a trigonal instability, joined by
a buckling of Si$_2$ planes, which is apparently\cite{Affronte} strongly
reduced under high pressure, or by Be substituting some of the Si sites.
The presence, or the proximity, of  structural instabilities
(investigated through phonon
spectra), will hopefully provide hints to further experimental search for new
materials.

In this paper, we present calculations
for \mgbtwo, \mgalbtwo, \bebtwo, \casitwo\ and \cabesitwo\ ($x=1$),
based on the local density approximation (LDA) to the density functional
theory.
We  use both  the Full Linearized Augmented Plane Wave
method, and the norm-conserving pseudopotential method.
In Sect.~\ref{sec:method} we give some computational detail; in
Sect.~\ref{sec:elec} and \ref{sec:casi} we discuss the electronic properties for all the
compounds
considered; we  finally draw our conclusion in Sect.~\ref{sec:concl}.

\section{Computational and structural details}
\label{sec:method}

Our calculations were performed using the ``all--electron"
full potential linearized augmented plane waves (FLAPW)
\cite{FLAPW}  method, in the local density approximation (LDA) to the
density
functional theory.
In the interstitial region we used plane waves with wave vector up to
$K_{max}$ = 3.9 a.u. and 4.1 a.u. for disilicides and diborides,
respectively (we used the larger $K_{max}$ = 4.4 a.u. for \bebtwo\ because of
the smaller spheres imposed by the smaller lattice parameters). 
Inside the  muffin tin spheres, we used an  angular momentum  expansion
up to $l_{max}$ = 6 for the potential and charge density and $l_{max}$ =
8 for the wave functions.
The Brillouin zone sampling was performed using the special $k$--points
technique according to  the Monkhorst-Pack scheme\cite{MP}, and also
the linear tetrahedron method with up to 120 $k$--points in the irreducible
Brillouin zone.
We used   muffin tin radii $R_{MT}$ = 2.2, 1.8,  1.59 and 2.1 a.u.,
for Ca, Si, B, and the remaining elements respectively (we used $R_{Be}$ = 2
a.u. and $R_{B}$ = 1.5 a.u. in \bebtwo).

The crystallographic structure of ${\rm AlB_2}$ (C32) is hexagonal,
space group $P6{mmm}$ with graphitic B   planes, and 12-fold
B--coordinated Al atoms sitting at the center  of B hexagonal prisms.
The unit cell contains one formula unit, with Al atom at the origin of the
coordinates and two
B atoms in the positions (${\bf d}_1=(a/2,a/2\sqrt{3},c/2)$,
${\bf d}_2=-{\bf d}_1$). The trigonal distortion in \casitwo\ corresponds to
a buckling of Si planes  leading to a $P3\overline{m}1$ space group symmetry), 
with one internal structural parameter $z$ (the internal coordinate of B atoms).

To calculate plasma frequencies, Fermi velocities,  Hall coefficients, and the
densities of states (DOS) near $E_F$,
we  have performed a  spline fit\cite{KW} of the {\em ab-initio} energy bands.
The fitted bands have then been computed over 128$^3$ points in the reciprocal 
space unit cell. The linear tetrahedron method has 
been used to compute DOS and the other quantities.

Phonon frequencies were computed using the ABINIT code~\cite{abinit}
by means of Linear Response Theory~\cite{bgt,gonze} (LRT)
approach in the framework of pseudopotentials plane wave
method\cite{abinit}.
Troullier and Martins\cite{TM} soft norm-conserving pseudopotentials
were used for all the elements considered, Mg and Ca $p$ semicore states
were not included in the valence, their contribution to the exchange and
correlation potential was accounted by the non-linear core corrections
~\cite{nlcc}.
The local density approximation to the exchange-correlation functional is
used.
Plane wave cutoffs of 44 Ry and 30 Ry were found sufficient to converge
structural and electronic 
properties of the diborides and silicides respectively. 
Brillouin zone integration was performed using a $20\times20\times20$
Monkhorst-Pack\cite{MP} mesh for the structural optimization and a 
$12\times12\times12$ mesh was used in the LRT selfconsistent calculation of 
phonon frequencies. 
To improve $k$--point sampling convergence
a gaussian smearing with an electronic temperature of 0.005 Ha was used.
The structure used in the LRT calculations was obtained optimizing 
the lattice structure up to a residual stress of 1 MPa.     

\section{Diborides}
\label{sec:elec}

\subsection{\mgbtwo\ and \bebtwo.}
\label{sec:puri}

 We shall start our discussion of electronic properties of diborides
 by presenting in Fig.\ \ref{bande_mg} the energy bands of the
superconducting
compound \mgbtwo, which will be the reference for all our further studies
(the $M-\Gamma-K$ lines are in the basal plane, while
the $L-A-H$ lines are the corresponding ones on the top plane at
$k_z=\pi/c$). We used the experimental lattice parameters
$a=3.083$ \AA, $ c=3.52 $ \AA.
Our results, which  are in excellent agreement with  those of
Kortus et. al.\cite{kortus}, show strong similarities with the band
structures
of most  simple-metal diborides\cite{mass}, computed by the FLAPW method.
 They   show  strongly bonded $sp^2$ hybrids laying in the horizontal
hexagonal
 planes,
forming the three lowest ($\sigma$--bonding)  bands,
as the main structure in the valence region. The corresponding antibonding
combinations are located around 6 eV above the Fermi level ($E_F$). 

The bonding along the vertical direction is provided by the $\pi$ bands
(at $\sim$ -3 eV  and $\sim$ 2 at $\Gamma$), forming the double  bell-shaped
structures.
Similarly to \albtwo-type diborides, and unlike graphite, we notice a large
dispersion
of the $\pi$ bands along the $k_z$ direction ($\Gamma-A$).
This can be understood on a tight-binding approach:  the phases
of the Bloch functions  lead the B $p_z$ orbitals on adjacent layers to
be antibonding at $\Gamma$, and bonding at $A$.
The downwards dispersion of
the $\pi$-bonding band (from $-3$ to $-7$ eV along $\Gamma-A$)
is joined by an opposite upwards dispersion of an
empty band, (at $\approx 2$~eV at the $\Gamma$ point).  
The analysis of the  wavefunctions for this empty
band reveals that this band has an interstitial
character, and is very similar  to those found in graphite\cite{posternak}, in
diborides\cite{mass}, in CaGa$_2$\cite{mass1}, but also in \casitwo.
A general description of the bonding can be given by the partial density of
states (PDOS), shown in Fig.~\ref{dosmg}. The B  PDOS, in particular, shows
 bonding and antibonding structures for the $s$ and $p$ states. The $p_z$
states give rise to a rather wide structure responsible for less than
half of the DOS at $E_F$, which  therefore results having predominantly
B $p-\sigma$ character.

In the search of similar materials having the desired superconducting
properties,
 \bebtwo\ is the first natural candidate, as the band filling is expected to
be pretty similar, and lighter Be atoms may help  providing larger phonon
frequencies while keeping similar electronic properties. The first 
experimental report on \bebtwo\cite{actacrys} gives the average lattice
constants. These experimental 
parameters have been used for all the previous calculations
of the \bebtwo\cite{tupitsin,mass}. We have optimized  the lattice
constants, imposing the \albtwo\ structure, obtaining $a=2.87$\AA, $c= 2.85$\AA,
smaller than in \mgbtwo\  and similar to the values estimated by the average
experimental values.
In particular, we notice that  while $a$ reduces only by about 7.4\%,
$c$ reduces by about 24\% when Mg is substituted by Be, a result which
can be understood pretty easily. On  one hand, in fact, the
optimization of $\sigma$-bonds  prevents a too drastic variation of the $a$
parameter;
 on the other hand, $c$ can change
more easily as the vertical bonding,  provided by the $\pi$ bands,
has a large contribution from the metal orbitals.
The band structure of \bebtwo, shown in Fig~\ref{bande_beb2}, are pretty
similar to
those computed by the FLAPW method\cite{mass} at the experimentally 
estimated lattice parameters.
We can remark the strong similarities with the \mgbtwo \ compound, 
which includes the presence of
$\sigma$-bonding hole
pockets along $\Gamma-A$, but also a few relevant differences.
First of all, the shorter lattice parameters lead to  wider valence bands,
and also to more dispersed $\sigma$ bands, in particular the cylindrical
hole pocket along $\Gamma-A$, which might be relevant for superconductivity.
Also, the different energy location of the metal $s$--free electron band
(lower in \mgbtwo), and the different shape        of the $\pi$-bonding band
(related by the different $c$ values)  causes a different occupation of this
band, especially   along $\Gamma-M$.

To further investigate  these two materials, and their relation to isostructural
diborides, we studied the zone center phonon frequencies of \mgbtwo,  \bebtwo, 
and \albtwo, using the linear response approach, with the pseudopotential
method. The calculations have been carried out at the theoretical lattice 
constants (also listed in Table \ref{phonons}), as obtained using the same method. 
 The two lowest independent 
frequencies correspond  to a motion of the intercalated metal atoms
relative to the rigid B networks: one vertical mode ($A_{2u}$),
 and one doubly degenerate in-plane mode, ($E_{1u}$). 
The two higher frequencies  represent the internal motion of the B network
itself: again  one vertical mode ($B_{1g}$), now representing a trigonal
distortion producing a buckling of the planes similar to that found in \casitwo,
and one degenerate ($E_{2g}$) mode giving an in-plane stretching of B-B bonds.
Our calculations for \mgbtwo \ are in good with other results
\cite{kortus,jepsenbohnen}.   
The comparison between the \mgbtwo\ and \bebtwo\ shows a much higher $E_{2g}$ frequency
in \bebtwo.  This result cannot be explained  through the different atomic
masses, as it involves only B motion.  It is therefore related to the different 
coupling of the electrons with the nuclear system.
Surprisingly,  two of the remaining frequencies are lower in
\bebtwo, namely the $A_{2u}$ mode which involves the motion of 
lighter Be atoms against the B network and the $B_{1g}$ mode.  
This indicates a strong
dependence of the phonon spectra of these materials on the electronic structure 
details. The lowest $E_{1u}$ mode has comparable frequencies. 
If we now go to \albtwo, we see again that the $E_{2g}$ mode is much higher
than in \mgbtwo.  Therefore, this mode has his minimum value in the
superconducting material. The modes involving the relative motion of the B$_2$
planes as a whole have frequencies comparable to the other materials; 
the $B_{1g}$, on the other hand, is smaller than in \mgbtwo\ and \bebtwo, which
is probably a consequence of the larger band filling (see the discussion on
\casitwo\ below).

We have computed, within a rigid band scheme and using the scheme described,
e.g., in Ref.\onlinecite{KW}, the Fermi velocities, plasma frequencies
(along the principal axes of the crystal),
and the independent components of the Hall tensor for \bebtwo\ and \mgbtwo.
Because of hexagonal symmetry, the non-zero components will be $R_{xyz}$,
corresponding to  the magnetic field  along the $z$ axis and transport
in plane, and $R_{zxy}=R_{yzx}$ corresponding to in--plane magnetic field.
The results, plotted in Fig.~\ref{trabemg} as a function of the hole or
electron-type doping, show relatively smooth variations over a wide range of
doping,
and are nearly coinciding, at zero doping,
with the corresponding calculations by Kortus
et al.\cite{kortus}.  After comparing  the two materials, the  following
remarks can be done:
(i) the  density of states at $E_F$ is smaller in \bebtwo.
(ii) \mgbtwo\ is nearly isotropic in terms of plasma frequencies and
Fermi velocities, while \bebtwo\ is expected to show important
anisotropy in resistivity, and, in particular, the in-plane conductivity is
expected to be smaller. A band-by-band decomposition explains these
results in term of very similar (and very anisotropic) contributions in
the two compounds from $\sigma$ states, and quite different contributions
from the $\pi$ bands, nearly isotropic in \mgbtwo\ and favoring conductivity
along
 the $c$-axis in \bebtwo.
  These results point out that despite structural similarity, the
electronic properties of \albtwo-type structures differ substantially from
graphite and intercalation graphite compounds.
(iii) The Hall coefficients are positive
(hole-like) when the magnetic field is along the $z$ axis  (transport
in--plane), while is negative for \mgbtwo\ and almost zero but negative for
\bebtwo, when the magnetic field is in--plane. A detailed analysis shows
again comparable contributions  in
the two compounds from $\sigma$  bands, while the sign of $R_{zxy}$ and
$R_{yzx}$ comes from a balance of the $\pi$  bands.
The Hall coefficient of  \mgbtwo\ has been recently measured
by Kang et al.\cite{hallexp}.  $R_H$, which will correspond in this case
to an average of the tensor components, turns out to be positive and to decrease
with temperature.  At $T=100$K they give 
$R_H= 4.1 \times 10^{-11}\rm m^3/C$, and an
extrapolation of their results at $T\rightarrow 0$,
gives $R_H\approx 6.5 \times 10^{-11}\rm m^3/C$. If we average, at zero doping,
our tensor components, the positive contributions  overcome  the  negative ones,
resulting in an average  value $R_H\approx 2 \times 10^{-11}\rm m^3/C$.  This
value is thus of the correct order, but single crystal measurements 
are probably necessary to have  a significant comparison, avoiding an
average of quantities with  different sign.  It is worth noticing that,
if the sign of carriers has a role on  superconductivity\cite{hirsch}, \bebtwo\
 also has a positive Hall coefficient.

\subsection{Al doping in \mgbtwo.}
\label{sec:doping}

A  recent experimental report by Slusky et al.\cite{cava}  shows  
that Al doping destroys bulk superconductivity when the Al content $x$ is
greater than $\approx 0.3$. Stimulated by these results, we  studied 
the chemical substitution of Mg with Al.
The  substitution corresponding to a 50\% concentration of Al ($x=0.5$)
leads to a reduction of the cell parameters,
 from $a=3.083$~\AA, $c=3.521$~\AA\cite{database_almg}  for \mgbtwo,
to $a=3.047$~\AA, $c=3.366$~\AA\cite{database_almg} for \mgalbtwo.
These differences again
indicate hardly compressible B-B $\sigma$ bonds, and interlayer distances
quite sensitive to the intercalated cation. The study of Slusky et
al.\cite{cava} shows
 consistent results, and  indicates, furthermore,  that after a 
two-phases region for Al
substitutions from 10 to 25\%, $c$ collapses and bulk superconductivity
disappears.  We first studied this problem
in a rigid band approach, bringing \mgbtwo\ to 
the same lattice parameters of $\rm Al_{0.3}Mg_{0.7}B_2$\cite{cava}.  The
only apparent change  at that electron concentration is the complete
filling of the $\pi$ bonding band at the $M$ point. The $\sigma$ bonding
bands need a  larger electron addition to be completely filled in a rigid
band model. 

 To investigate further this problem, we studied the
system with 50\% Al concentration.
We used in our calculations the experimental values, and in a first approach
simulated the disorder (certainly present in the real compound) using an
orthorhombic supercell. The corresponding unit cell contains
two formula units, and is described by the two orthogonal in plane basis vectors 
 ${\bf a}_1=(a,0,0)$ and  ${\bf a}_2=(0,a \sqrt {3},0) $.
We started by calculating the energy bands using the \mgbtwo\ lattice
constants, and then we used the experimental \mgalbtwo\ constants.
The results of these  calculations are shown in Fig.~\ref{mgal}
together with the bands of \mgbtwo, folded into the Brillouin zone (BZ) of the
orthorhombic structure. The folding is such that the $\Gamma$ point of the
orthorhombic BZ corresponds to
the $\Gamma$ and $M$ points of the hexagonal BZ, while $Z$ corresponds to
both the hexagonal
$A$ and $L$.  Therefore, the $\Gamma-Z$ line collects the $\Gamma-A$ and the
$M-L$ hexagonal lines, and we notice immediately the three $\pi$ bands (two
bonding and one anti-bonding, coming from folding). The $\Gamma-X$ line
contains the (folded) hexagonal $\Gamma-M$ line.
If we  first look at the frozen structure calculation Fig.~\ref{mgal}(c), we
see
quite similar band dispersions with the obvious splittings,
related to the symmetry lowering induced by the  different cations, and a
different band filling. We notice, however, a lowering  of the folded $\pi$
antibonding band,
related to the different cations:  the $\pi$ band is in fact pushed downwards by
the interaction with the free-electron  metal-$s$  band, which is at 
lower energy in the
compounds with Al (this band is partially filled in pure, non superconducting,
\albtwo\cite{mass}). As a consequence,
the extra electrons coming from Al do not fill the $\sigma$-bonding band
completely, leaving
hole pockets around $Z$. As we relax the structure,
the $\sigma$ bands increase their
width due to the smaller $a$ value, and the situation near $E_F$ changes
only because
of the occupation of the free-electron band around $\Gamma$.
These results suggest that although  the rigid
band scheme can be considered roughly correct, the  dependence  upon doping 
of the physical properties and  of Fermi surface states relevant for
superconductivity, may be crucially dependent on the actual substitutional
atom.

\section{\casitwo\ and related compounds}
\label{sec:casi}

When prepared at  ambient pressure $\rm CaSi_2$ has a rhombohedral
structure (space group $R3\overline{m}$, $a=3.85$ \AA, $c=30.62$ \AA)
\cite{Pearson} and is a semimetal \cite{Fahy},
not superconducting down to $30$~mK\cite{Affronte2}.
By annealing under pressure - typical conditions are $8$~GPa and
$800$~K - a tetragonal superconducting
($\rm \alpha-ThSi2-type$) phase appears with a 3-dimensional network
\cite{Evers}  of Si, showing superconducivity with  $T_c=1.58$~K.
At pressures between $\approx 7$~GPa and $9.5$~GPa, rhombohedral structure
samples undergo a transition to a trigonal phase ($P3\overline{m}1$,
$a_T \approx 3.78$~\AA, $c_T \approx c_R/6\approx 4.59$~\AA) with
 Si at  $2d$ positions,  and   $z=0.4$, where $z$ is the coordinate of the 
B atoms in units of $c$ ($z=0.5$ corresponds to the perfect \albtwo\ structure).
A second phase transition occurs at $16$~GPa. This new phase may be
described as \cite{Affronte} an $\rm AlB_2$-like structure  with $z$ slightly
smaller than the ideal value $1/2$.
The ${\rm AlB_2}$-type structure was previously observed in rare-earth
 silicides \cite{database_disil}, while in \casitwo\ it is apparently
stabilized by doping the Si planes with Be\cite{database_cabesi}. 

  In Fig.  \ref{bande_casi2} we show the
energy bands  of \casitwo\ in the \albtwo\ structure (dashed lines),
superimposed with those of the
trigonal structure (full lines), along the main symmetry
lines of the Brillouin zone.
Since this polymorph can be obtained only under pressure, we do not minimize
the lattice parameters, but rather use  the experimental values 
$a=3.7077$~\AA\ and $c=4.0277$~\AA,  corresponding to P=19.3 GPa,
and optimize the internal parameter $z$. The bands in the \albtwo\ structure
show strong similarities with  those of  simple metal diborides, and of
CaGa$_2$\cite{mass}. The major difference is provided by Ca $d$ states,
which are mostly found in the
conduction region, above 4 eV;  however, there is a
{\it p-d} mixing between Si and Ca, especially concerning Si $\pi$ and Ca
$d_{z^2}$ orbitals, also mixing with the free-electron states.
Unlike  in \mgbtwo, the  Fermi level of \casitwo\ cuts only the antibonding 
part of the $\pi$ bands, while the  bonding part is fully occupied. 
This of course implies a reduced
contribution of $\pi$ states to the stabilization of an $sp^2$ environment,
consistent with the $sp^3$-like trigonal distortion.  We will come back to
this point later. The energy bands in the trigonal structure show that, a
part from a small rigid shift, the $\sigma$ bands change very little
with respect to the \albtwo\ structure.  The
$\pi$   bands, on the other hand, show anticrossings which in the
antibonding
case are located right at the Fermi level.
This is most clearly indicated in the density of states of \casitwo,
reported
in Fig.~\ref{doscasi2}. The dip at $E_F$ in the trigonal structure
provides a textbook explanation for the lattice distortion.  

In order to understand the stability of $\rm AlB_2$-type polymorph,
 we study the total energy
variation as a function of $z$, relative to the ideal $z=0.5$ case, for two
experimental  structures\cite{Affronte}, and report it in
Fig.~\ref{etot}(a). There is a good agreement between theory and experiment
(within 2.5\%), 
and a non-negligible ($\approx 30$ mRy per formula unit) stabilization energy.
We notice, however, that the stabilization energy decreases with pressure.
As we fix $a$ and $c$  to the values corresponding to the larger
 pressure of 19.3 GPa\cite{Affronte},  we see 
a larger discrepancy relative to the experimental equilibrium value of
 $z$  ($z_{exp}\sim 0.448$). This
 indicates that some different modification might be
going on in the experimental samples. 
In order to discuss the link between the stability of \casitwo\ in the 
\albtwo\ structure and the $\pi-$band filling, 
without modifying the Si$_2$ planes themselves, 
we vary the atomic number of the cation, which now represents a virtual
atom ranging from Ca to K (corresponding to $Z=20+x$). Our results,
reported in Fig~\ref{etot}(b), show that while $x=-0.3$ leaves the total
energy curve almost unchanged, $x= -0.7$ or larger give a nearly vanishing
trigonal stabilization energy. This is consistent with our previous
conclusion that a filling of $\pi$-antibonding bands destabilizes the
\albtwo\ structure.

This last conclusion is supported by the existence, in the hexagonal
structure, of \cabesitwo, with $x=0.75$, at $a=3.94$~\AA\ and $c=4.38$~\AA
\cite{database_cabesi}.
In fact, Be doping removes
electrons from the Si$_2$ planes, and brings the formal filling of the above
$\pi$-antibonding bands to the level found in diborides. To investigate
these effects, we studied the (artificially ordered, but existing)
\cabesitwo\ system
for  $x=1$, not far from the experimental stoichiometry. The corresponding
bands,  shown in Fig.~\ref{bands_cabesi},
 have been calculated at the experimental
lattice parameters ordering Be and Si atoms as, e.g., in
hexagonal BN. In this way, the Brillouin zone is the same as
in the undoped compound, and the interpretation is easier.
The CaBeSi band structure shown are overall
similar to that of the undoped compound. The symmetry lowering induced by
substitution produces the large splitting
of degeneracies at the  K and H points.   The major difference is of course
related to the position of the Fermi level, which is almost exactly
coinciding with that found in the superconducting \mgbtwo.
The reduced bandwidth for the $\sigma$ bonding bands may be associated with
the smaller size of Be orbitals.
The electron depletion from the antibonding $\pi$-bands removes
the trigonal instability, as demonstrated by total energy calculations. 
In fact, while the $\Gamma-$point phonons of \casitwo\ in the \albtwo\ structure
show the presence of an instable phonon, in agreement with FLAPW total energy 
calculations, in \cabesi\ the phonons are stable, as shown in 
Table \ref{phonons}. Table \ref{phonons} shows that this system has lower 
phonon frequencies than the diborides, as is normal given the heavier atomic
species involved. If we now immagine to vary $\rm x$ in \cabesitwo, we will 
therefore have that, at some value of $x$, the $B_{1g}$ mode,  
corresponding to the trigonal distortion, will 
be close to an instability. This situation  may thus offer an
interesting playground in the search for superconducting materials.

The Fermi level of \cabesi\ cuts the $\sigma$ bands, leaving hole pockets quite
similar to those of \mgbtwo. 
While it is  difficult to say, on the basis of our electronic structure
calculations only, whether the similarities with \mgbtwo\
band structure can lead to superconductivity, it may be useful to provide
hints on
the stability of this compound, which may help experimentalists in the
search for
new compounds.  The PDOS in Fig.~\ref{dos_cabesi} show a strong
hybridization
between Si and Be states, in the valence region.  In other
words, substitutions on the Si site can be possible without disrupting the
$sp^2$ network if the one-electron energies of the impurity atoms are 
comparable with
those of Si, and light atoms should be preferred in order to keep high
values of
phonon frequencies. We may suggest B, which however corresponds to a hole
doping
smaller than Be.

\section{Conclusions}
\label{sec:concl}

We have performed electronic structure calculations for \mgbtwo,
\bebtwo, \mgalbtwo,  $\rm CaSi_2$ in the trigonal and in the
  $\rm AlB_2$-type polymorphs at the high pressure experimental
structural parameters, and for the
$\rm CaBe_xSi_{2-x}$ system ($x=1$), experimentally stabilized for $x=0.75$.
We have calculated band structures,  PDOS, transport
properties and phonon frequencies, using the FLAPW and the norm-conserving
ab-initio pseudopotential methods, within the local density
approximation.  The following conclusions can be drawn on the basis of our
calculations:  as compared to  \mgbtwo, \bebtwo\ has a similar filling of 
$\sigma$ bands, but differs
in terms of $\pi$ states near $E_F$. The DOS at $E_F$ is lower, and some phonon
frequencies are substantialy higher while others are comparable or lower.  
It would be interesting to further investigate  this compound. 
Al substitution cannot be fully simulated by a rigid band model, as the
$\pi$ antibonding states appear to be quite sensitive to Al substitution,
and absorb much
of the added electrons. Transport properties show a substantial
two-dimensionality of
these compounds,  due to $\pi$ states.

We  studied \casitwo\ in its trigonal structure, and show that the
distortion can be removed only by subtracting (in a virtual crystal-like
approximation) at least $0.7$ electrons.
In fact, \cabesi\ shows no tendency towards  trigonal distortion, and
electron states near $E_F$ of similar bonding nature to those of \mgbtwo.

\section{Acknowledgements}
We thank M. Affronte    and A. Gauzzi 
for stimulating discussions, and M.Affronte for sharing his
results prior to publication.
This work was partially supported by the italian Consiglio 
Nazionale delle Ricerche (CNR) through the 
``Progetto 5\% Applicazioni della superconduttivit\`a ad alta T$_c$".

\begin{table}
\centering
\small
\caption{ Theoretical structural parameters (in \AA) and phonon frequencies 
at $\Gamma$  (in $\rm cm^{-1}$) for the \albtwo\-type systems. 
Level degenaracies are shown in parenthesis.}
\vspace{5mm}
\begin{tabular}{|c|c|c|c|c|}  
 \hline 
             & MgB2 &  BeB2   & AlB2 &  CaBeSi \\
 \hline \hline
a            & 3.045 & 2.903  & 2.976&  3.914  \\
c            & 3.480 & 2.853  & 3.237&  4.488  \\
(2) $E_{1u}$ & 349  &  327    & 306  &  189   \\
(1) $A_{2u}$ & 426  &  213    & 437  &  247   \\
(2) $E_{2g}$ & 575  &  893    & 996  &  461   \\
(1) $B_{1g}$ & 708  &  642    & 515  &  346   \\
 \hline
\end{tabular}
\label{phonons}
\end{table}

\begin{figure}
\caption{Band structure
of $\rm MgB_2$. All the energies are referred to the Fermi level, taken
as zero.}
\label{bande_mg}
\end{figure}

\begin{figure}
 \caption{Total and partial density of states (PDOS) for $\rm MgB_2$.
}
\label{dosmg}
\end{figure}

\begin{figure}
\caption{Band structure
of $\rm BeB_2$.}
\label{bande_beb2}
\end{figure}

\begin{figure}
\caption{ Transport properties of \mgbtwo\ and \bebtwo\ as a function of
doping in a rigid band scheme. Full and dashed lines correspond to
\bebtwo\ and \mgbtwo\ respectively. Circles  (squares) indicate
 the $x,y$  ($z$) principal
values  of the plasma frequency $\Omega_p$ (in eV), of the mean
squared Fermi velocity $v_F$ (in $10^7$~cm/sec) and of  $R_{xyz}$  (in
$10^{-10}\; \rm m^3/C$, notice that $R_{zxy}=R_{yzx}$). DOS are in
states/eV-cell.}
\label{trabemg}
\end{figure}

\begin{figure}
\caption{Band structure of $\rm MgB_2$ (panel (a))  and  \mgalbtwo\
at the experimental (panel (b)) and frozen $\rm MgB_2$ (panel(c)) lattice
parameters.}

\label{mgal}
\end{figure}

\begin{figure}
\caption{Band structure
of $\rm CaSi_2$ in the ideal \albtwo\ (dashed lines) and  trigonal
(full  lines) structures .}
\label{bande_casi2}
\end{figure}

\begin{figure}
\caption{
Total and partial density of states (PDOS) for $\rm CaSi_2$ in the \albtwo\
 (dashed lines) and in the trigonal
structure (full  lines).
}
\label{doscasi2}
\end{figure}

\begin{figure}
\caption{ Upper pannel:
total energy  of $\rm CaSi_2$
as a function of the internal parameter $z$, for two different values
of pressure, at the corresponding experimental lattice constants 
( $a=3.7554$~\AA, $c=4.3951$~\AA\ for $P=15.0$~GPa and 
$a=3.7668$~\AA, $c=4.4752$~\AA\ for $P=12.8$~GPa, after
\protect \cite{Affronte}).  The experimental values of $z$ are indicated 
by arrows.
Lower pannel: same as above, at the larger experimental
pressure (19.3 GPa),  with Ca substituted by a
virtual cation having nuclear charge $Z=20+x$.  The experimental lattice
parameters  are $a=3.708$\AA\ and $c=4.028$\AA,  $z=0.448$).
The zero of energy corresponds to the \albtwo\ value $z=0.5$.}
\label{etot}
\end{figure}

\begin{figure}
\caption{Band structure of $\rm CaBe_xSi_{2-x}$ ($x=1$) 
in the hexagonal Brillouin zone,  at experimental
lattice parameters.}
\label{bands_cabesi}
\end{figure}

\begin{figure}
\caption{
Total and partial density of states (PDOS) for $\rm CaBe_xSi_{2-x}$ ($x=1$)
in the \albtwo\ structures (see text).}
\label{dos_cabesi}
\end{figure}

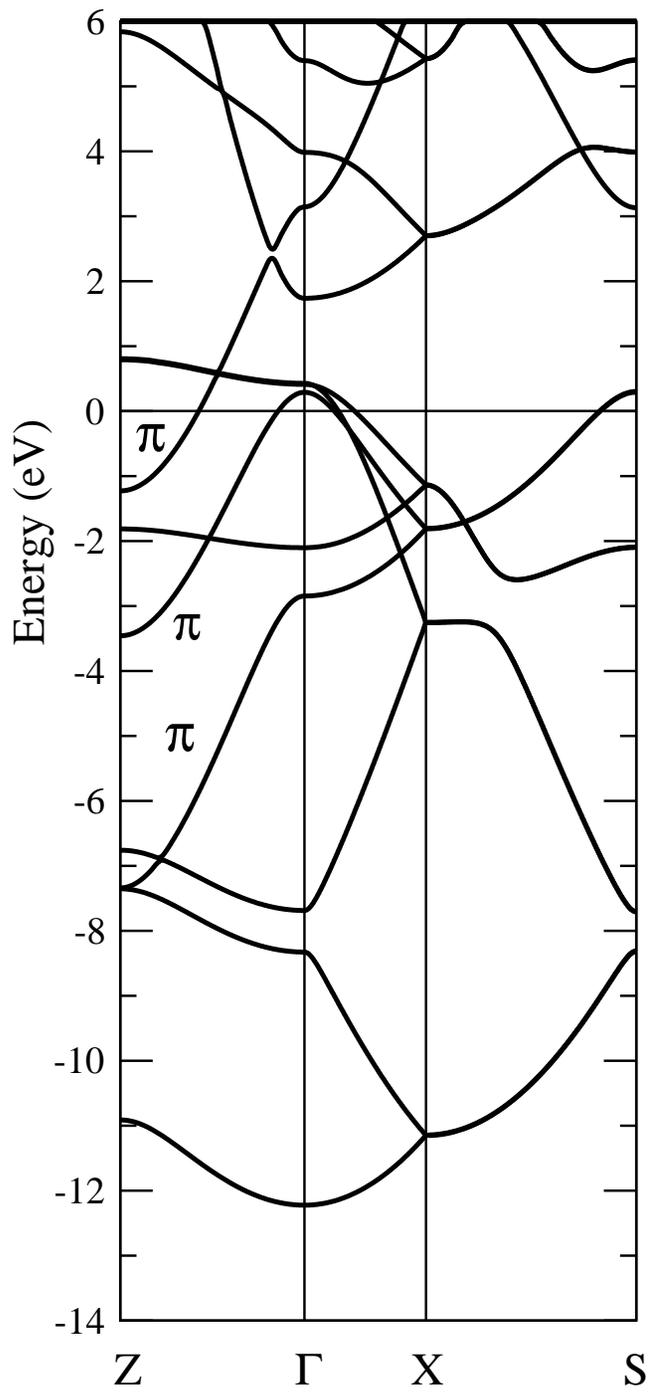
\end{document}